\documentclass[12pt]{article}
\usepackage{latexsym,amsfonts,amssymb,amsmath}
\usepackage[utf8]{inputenc}

\newcommand{\ri}{{\mathrm i}}
\newcommand{\p}{\partial}
\newcommand{\bea}{\begin{array}}
\newcommand{\eea}{\end{array}}

\catcode`@=11 \long
\def\@caption#1[#2]#3{\par\addcontentsline{\csname
ext@#1\endcsname}{#1} {\protect\numberline{\csname
the#1\endcsname}{\ignorespaces #2}} \begingroup \small
\@parboxrestore \@makecaption{\csname fnum@#1\endcsname}
{\ignorespaces #3}\par \endgroup} \catcode`@=12

\newcommand{\la}{\label}

\catcode`@=11 \long
\def\@caption#1[#2]#3{\par\addcontentsline{\csname
ext@#1\endcsname}{#1} {\protect\numberline{\csname
the#1\endcsname}{\ignorespaces #2}} \begingroup \small
\@parboxrestore \@makecaption{\csname fnum@#1\endcsname}
{\ignorespaces #3}\par \endgroup} \catcode`@=12

\begin{document}

\allowdisplaybreaks
 \begin{titlepage} \vskip 2cm

\begin{center} {\Large\bf Symmetries of  Schr\"odinger equation with scalar and vector potentials}
\footnote{E-mail: {\tt nikitin@imath.kiev.ua} } \vskip 3cm {\bf {A.
G. Nikitin } \vskip 5pt {\sl Institute of Mathematics, National
Academy of Sciences of Ukraine,\\ 3 Tereshchenkivs'ka Street,
Kyiv-4, Ukraine, 01601\\}}\end{center} \vskip .5cm \rm

\begin{abstract}Using the algebraic approach Lie symmetries of time dependent Schr\"odinger equations for charged particles interacting with   superpositions of scalar and vector potentials are classified. Namely, all the inequivalent equations admitting symmetry transformations with respect to continuous groups of transformations  are presented. This classification is completed and includes the specification of symmetries and  admissible equivalence relations for such equations. In particular, a simple mapping between the free Schr\"odinger equation and the repulsive oscillator is found which has a clear group-theoretical sense.
\end{abstract}
\end{titlepage}
\section{Introduction\label{int}}

    Symmetries are very important constituent parts of  modern physics. Especially important role are played by symmetries in quantum mechanics. In particular, the most fundamental systems of quantum mechanics like the Hydrogen atom or harmonic oscillator admit  extended number of symmetries, and may be just this property make them fundamental.

     There are different types of symmetries arising in quantum mechanics. In particular, we can mention the invariance of motion equations  w.r.t. continuous groups of transformations. Just symmetries of this type can be treated as basic ones. In particular  the invariance of the (free) Schr\"odinger equation (SE) with respect to  Galilei group is in a good accordance with the Galilei relativity principle.

     I addition, there is a lot of other important symmetries accepted by  SE. They are the higher symmetries which are nothing but integrals of motion belonging to differential operators of order higher than one. A perfect example of such symmetries is the Laplace-Runge-Lentz vector for the Hydrogen atom.

     The higher symmetries  give rise to such nice properties of SE as superintegrability and supersymmetry, see surveys \cite{PW} and \cite{NNN}. Let us remind that the inverse problem approach also is based on higher order integrals of motion. Just the higher symmetries made it possible to extend the number of conservation lows for the electromagnetic field \cite{Fu}.

    The search for symmetries of SE has a long and glorious history. In fact it started already in nineteenth century when Sophus Lie discovered the maximal symmetry group of  the linear heat equation. Since the SE is nothing but the complex form of the latter one, it keeps the symmetries discovered by Lie.

     These results were recovered and developed
in papers \cite{Nied}, \cite{And} and
\cite{Boy}.  Niederer \cite{Nied} had found the maximal invariance group of the free SE. In particular he  shows that in addition to the Galilei invariance, this equation admits the dilations and conformal transformations.

Symmetries of the one dimensional SE with a non-trivial potential were described in \cite{And}. Boyer \cite{Boy} had made the same business with   two- and three-dimensional systems.  These results   form  an important part  of the group-theoretical grounds of quantum mechanics. In particular, the a priori information on possible symmetries w.r.t.  the continuous groups of transformation is necessary for effective  description of the coordinate systems which generate solutions of SE in separated variables \cite{Mil}. It is important also for the investigation of higher symmetries. But maybe the main value of the Boyer results is that they contain the  information about all possible symmetry groups which can be accepted by the basic equation of quantum mechanics.

The complete description of 2d quantum mechanical systems admitting the second order integrals of motion was presented in \cite{Wint} and \cite{Wint1}. And it needed as much as twenty four years to extend this result to the case of the 3d systems, see papers \cite{evan, evan2}.

The search for higher  symmetries of SE and its generalizations (such as Schr\"odinger-Pauli equation and SE with position dependent mass (PDM)) is still a very popular business. The modern trend in this field is the study of  the third and even arbitrary order integrals of motion \cite{marqu, marq, snoba}  se also paper \cite{NikNik} where the determining equations for such symmetries were presented.

It happens that  the group classification of PDM SEs
 was waited for a very long time.
The complete group classification of
these equations appears  recently in papers \cite{NZ}  and \cite{NZ2,NN} for the stationary and time dependent
cases correspondingly. In paper \cite{N1} we start the systematic search for the higher order symmetries in the PDM systems.

Symmetries of the Schr\"odinger-Pauli equations with respect to the continuous groups of transformations are classified in \cite{NNNN}. But it had been done only for such equations which describe  chargeless  particles like the neutron.
Higher symmetries of such equations  have been studied   also. We can mention the papers devoted to its supersymmetries \cite{kar, N5},  higher order symmetries \cite{N6,N7} and  Fock symmetries \cite{N8,N9}. Relativistic systems with Pauli interaction which admit   the mentioned symmetries were studied in \cite{Mafia, beck1, beck2}.

But all the mentioned results are restricted to SEs with scalar or matrix potentials and have nothing to do with equations including the vector potential of the external electromagnetic  field. The higher symmetries of SE for charged particles interacting with magnetic fields is a relatively new research field, their systematic study starts recently with papers \cite{snobl, snobl1}. Let us mention that the problem of separation of variables in such equations was discussed  long time ago in paper \cite{Zha}.

However, a description of continuous  symmetries for SEs including superpositions of scalar and  vector  potentials is still missing, and this  is a strong challenge for experts in group theoretical methods in physics. Indeed, such equations are more general and  by no means less fundamental than SEs with scalar potentials, and the classification of their symmetries would be an important part of the group-theoretical grounds of quantum mechanics. And all the  arguments adduced  in the above to prove the significance of the group classification SEs equations with scalar potentials are valid for the case when the vector potentials are present also.  In addition, the symmetry classification of SEs with scalar and vector potentials is a rather complicated mathematical problem since such equations include as many as three arbitrary elements because the potentials should be treated as unknowns. To solve this problem the refined methods of the group analysis of  differential equations  are requested.

Just this problem is solved in the present paper. Namely, we give the complete classification of SEs with scalar and vector potentials which admit inequivalent symmetries.  To achieve this goal we use the so called algebraic approach whose the main idea is the a priory study of the admissible algebraic structures of symmetry operators which can be admitted by the classified differential equations. In particular, the group classification of the considered class of differential equations can be reduced  to the classification of  subalgebras of the associated equivalence algebra.

The algebraic approach presents very effective tools for solving the determining equations for coefficients of symmetry operators and make the classification results very reliable. Just this approach was used in the above cited papers \cite{NZ,NZ2,NN} and paper \cite{popa1} where symmetries of (1+1)--dimensional SEs with time dependent complex potentials where classified.
However, its application   can be recognized already in more old paper \cite{gan}.

An important integral part of the {\it complete} group classification of any class of PDEs  is the description of all admissible equivalence relations for this class, i.e., of such transformations of the dependent and independent variables which keep the generic form but can change the arbitrary elements of the classified equations. The equivalence transformations are requested for a clear description of inequivalent symmetries but also have their own application value.

 Thus, in addition to symmetries  of the SEs we were supposed  to describe the
 equivalence relations between certain classes of them.  We make our best to make such
description in a convincing and clear manner. In particular a new connection between the free SE and repulsive oscillator is presented which has the exact group-theoretical meaning and is seemed to be more elegant than the known ones. Moreover,   this connection is generalized to the extended classes of the classified equations which are invariant with respect to scaling of the independent variables.

Summarizing, we find all inequivalent versions of SEs with scalar and vector potentials which admit symmetries w.r.t. continuous transformation groups. This result supplies us by the important information about the properties of one of the basic equations of quantum mechanics. It can be used in mathematical modeling of physical systems with a priori requested symmetry properties, in searching for the models which admit solutions in separated variables, in construction of superintegrable and exactly solvable models, etc., etc. Finally, it is a certain contribution to our general understanding of these fundamental  equations.

 Let us describe the structure of the paper. In Introduction a short survey of the main results concerning symmetries of the (generalized) SEs is given.
In Section 2 the analysed equation and the main notations are presented. In Section 3 the determining equations for symmetries and their useful algebraic consequences are deduced. In Section 4 we apply the classical Lie result concerning continuous groups on a line to describe the time dependence of the symmetry group generators. Section 5 is devoted to the description of the equivalence groupoid for the considered class of equations. In Sections 6 and 7 the determining equations are solved and the classification results are presented. Finally, in Section 8 the obtained results are discussed in more detail.

\section{Main definitions}

Let us start with  the standard SE equation  for a charged particle interacting with the external electromagnetic field and write it in the following form:
\begin{gather}\left(\ri\frac{\p}{\p t}-H\right)\psi(t,{\bf x})=0\la{se}\end{gather}
The corresponding Hamiltonian $H$ looks as follows:
\begin{gather} \la{H1} H=\frac12\pi_a\pi_a+eA^0\end{gather}
where
\begin{gather}\la{pi} \pi_a=p_a-eA^a, \ p_a=-\ri\frac{\p}{\p{x_a}},\end{gather}
 $A^0$ and $A^a$ are components of the vector potential of the  electromagnetic field, $e$ is the  coupling constant associated with the particle charge, which, up to the redefinition of $A^0$ and $A^a$ can be (and will be) chosen as the unity,
 and summation is imposed over the repeating indices $a$ over the values $a=1, 2, 3$. Moreover, the potentials of the external field are supposed to be time independent.

The vector-potential should satisfy the continuity equation which, in view of its time independence, is reduced to the divergenceless condition for the vector potential. However we prefer to change it by the condition $A^3=0$ which always can be done using the gauge transformations. In particular  we can set:
\begin{gather}\la{vp} A^3=0,\quad A^1=\p_1F({\bf x})+\p_2G({\bf x}), \quad A^2=\p_2F_1({\bf x})-\p_1G({\bf x})\end{gather}
were $F_1=\frac{\p F}{\p x_1}$, etc,  $F$ and $G$ are functions of $\bf x$.
In this case the vector potential is not supposed to be divergenceless, and it is the price which we pay for the reduction of the number of arbitrary elements, i.e., non-trivial components of the potential.

To simplify the following calculations we  express functions $A^1$ and $A^2$ via another functions, i.e., $F$ and $G$. Notice that just function $G$ generates the divergenceless part of the vector-potential.

Substituting (\ref{pi}) and (\ref{vp}) into (\ref{H1}) we reduce $H$ to the following form:
\begin{gather}\la{H2}H=\frac12 p_ap_a-\frac{1}2(\{A^1,p_1\}+\{A^2,p_2\})+V\end{gather}
where
\begin{gather}\la{sp}V=A^0+\frac{1}2({(A^1)}^2+{(A^2)}^2)\end{gather}
 and $\{A^1,p_1\}=A^1p_1+p_1A^1$.

Equation (\ref{se}) with Hamiltonian (\ref{H2}) includes three arbitrary elements, i.e., functions  $V$,  $F$ and $G$.  Moreover, setting in (\ref{H2}) $A^1=A^2=0$  we come to the  standard SEs with a scalar potential $V$, which is a very particular  subject of our analysis.

\section{Determining equations}

We will search for symmetries of equations (\ref{se}) with respect
to continuous groups of transformations of dependent and independent variables. To do it we can  use the classical Lie algorithm whose contemporary version can
be found in \cite{olver}.   In the case of the linear SE this algorithm can be reduced to searching for
the first order differential operators of the following form:
\begin{gather}\label{so}
    Q=\xi^0\partial _t+\xi^a\partial_a+\tilde\eta\equiv \xi^0\partial _t+
    \frac12\left(\xi^a\partial_a+\partial_a\xi^a\right)+\ri\eta,
\end{gather}
where $\tilde
\eta=\frac12\xi^a_a+\ri\eta,\ \ $ $\xi^0$, $\xi^a$ and $\eta$ are
functions of $t,{\mathbf x} $.

Generator (\ref{so}) transforms solutions of equation (\ref{se}) into solutions if it satisfies the  following operator equation
\begin{gather}\la{ic}[Q,L]\equiv QL-LQ=\alpha L\end{gather}
where $L=\ri\p_t-H$ and $\alpha$ is one more unknown function of $t$ and $\bf x$.

Evaluating the commutator in the l.h.s. of (\ref{ic}) and equating coefficients for the linearly independent
differentials we obtain the following system of  determining equations:
\begin{gather}
\dot\xi^0=-a, \quad  \xi^0_a=0,\la{de1}\\
\la{de2} \xi^b_{a}+\xi^a_{b}-\frac{2}3\delta_{ab}\xi^i_i=0,
\\\label{de6} \xi^i_i=-\frac{3}2\alpha,\\ \eta_a=A^b\xi^a_b-\xi^bA^a_b+\alpha A^a-
\dot\xi^a
,\label{de7}\\
\label{de8}\xi^aV_a=\alpha V+\dot\eta
-eA^a\eta_a\end{gather}
where $\dot\eta=\frac{\p \eta}{\p t} $ and $\eta_a=\frac{\p \eta}{\p x_a}$, i.e., the dot and subindex $a$ denote the derivation w.r.t. $t$ and $x_a$ respectively.

The system of the determining  equations (\ref{de1})--(\ref{de8}) is rather complicated and includes tree arbitrary elements $ A^1, A^2 $ and $V$ whose form should be fixed by the compatibility condition of this system. However, the subsystem (\ref{de1}), (\ref{de2}), and (\ref{de6}) does not include these arbitrary elements. Its generic solution is well known and can be represented in the following form:
\begin{gather}\la{kil} \xi^a=
-\frac\alpha2 x_a+\theta^{ab} x_b+\nu_a,\ \alpha=-\dot \xi^0\end{gather}
where $\xi^0, \alpha$,  and $\nu^a$ are functions of $t$ and $\theta^{ab}=-\theta^{ba}$ are constants.

Substituting (\ref{kil}) into (\ref{de7}) and integrating with respect to $x_a$ we obtain the following generic form of $\eta$:
\begin{gather}\la{eta1}\eta^0=\frac{\dot\alpha}4x^2-\dot\nu^ax_a+
K({\bf x},t)=0\end{gather}
where parameters $\alpha$ and $\nu$ should satisfy the condition $\alpha\nu^a=0$. In addition,
$K({\bf x},t)$ is a yet  unknown function which includes both the integration constant (which can depend on $t$) and the formal integral of the terms including the vector potential. In accordance with (\ref{de7}) and (\ref{eta1})
$K({\bf x},t)$  satisfies the following condition:
\begin{gather}\la{con10} K_a=A^b\xi^a_b-\xi^bA^a_b+\alpha
A^a
\end{gather}
which, in view of (\ref{kil}), can be reduced to the following form:
\begin{gather}K_a=\frac\alpha2\left(
A^a_bx_b+A^a\right)+\theta^{ab}A^b-\theta^{bc}x_cA^a_b-\nu^bA^a_b.
\la{eqA}\end{gather}

One more equation for  potentials  is given by (\ref{de8}). Using again relation (\ref{kil}) and taking into account the time independence of $\theta^{ab}$ we transform it to the form:
\begin{gather}\la{con10b}\left(-\frac{\alpha}2x^a+\theta^{ab}x_b+\nu^a\right)A^0_a+\frac{\dot \alpha}2x^aA^a-\dot\nu^aA^a=\alpha A^0+\frac{\ddot \alpha}2 r^2-\ddot\nu^ax^a+\dot K.\end{gather}

In addition to (\ref{eqA}) it is convenient to consider  the  following  algebraic consequences:
\begin{gather}\la{con12}\frac{a}2x_a(x\cdot A)_a-\theta^{dc}x_c(x\cdot A)_d-\nu^b(x\cdot A)_b+\nu^bA^b=x_aK_a,\end{gather}
\begin{gather}\la{con13}\frac{a}2x_a(x\times A)^n_a-\theta^{dc}x_c(x\times A)^n_d-\nu^b(x\times A)_b^n+(\nu\times A)^n=(x\times \p)^n K\end{gather}
where $x\cdot A=x_aA^a,\ (x\times A)^n=\varepsilon^{nab}x_aA^b$, etc, with $\varepsilon^{nbc}$ being the absolutely antisymmetric unit tensor. The system of equations  (\ref{con12}) and  (\ref{con13}) is equivalent to
(\ref{eqA}).

Thus to  classify symmetries of the SE which include a superposition of vector and scalar potentials, i.e., of equation (\ref{se})   it is necessary to solve the system of equations (\ref{eqA}) and (\ref{con10b}). This system includes parameters $\alpha$ and $\nu$ which can depend on time, and  time independent parameters $\theta^{ab}$.
\section{Dependence of symmetries on time}
Let us specify the possible time dependence of the parameters involved into  equations (\ref{eqA}) and (\ref{con10b}). To do it we return to equation (\ref{so}) defining the generic  form of symmetry operators. In accordance with conditions (\ref{de1}) function $\xi^0$ does not depend on $\bf x$, and so our symmetries include the time derivative term of the following form:
\begin{gather}\la{t1}Q=\xi^0(t)\p_t.\end{gather}

Operators  (\ref{so}) should form a basis of a Lie algebra, the same is true for their part presented in (\ref{t1}). The latter one represents generators of Lie groups on a line, and it is
known since time immemorial
that such groups are as maximum three parametrical, and so the corresponding Lie algebra is three dimensional.  Moreover, up to a change of variables these generators can be written as:
\begin{gather}\la{t2}Q_1=\p_t,\quad Q_2=t\p_t,\quad Q_3=t^2\p_t.\end{gather}
The corresponding function $\xi^0$ in (\ref{so}) is constant, linear or quadratic. In addition, generators
(\ref{t2}) form a basis of algebra $\textsf{sl}(2,R)$.

Changing variables in (\ref{t2}) we can obtain another realizations of this algebra whose generic form is:
\begin{gather}\la{t3}\tilde Q_1=f(\tilde t)\p_{\tilde t},\ \tilde Q_2=g(\tilde t)\p_{\tilde t},\ \tilde Q_3=h(\tilde t)\p_{\tilde t}\end{gather}
where $f(\tilde t), g(\tilde t)$ and $h(\tilde t)$ are functions of the changed variable $\tilde t=F(t)$.
 However, such changes are not necessary compatible with the symmetries of SE. To find the admissible changes of the time variable we use the following fact:   up to choosing  a basis one of operators (\ref{t3}), say $\tilde Q_3$,  has to have a constant multiplier for $\p_{\tilde t}$, i.e., function $h(\tilde t)$ should be a constant which can be normalized to the unity. Indeed, such operator is necessarily present in the list of symmetries of equation (\ref{se}).

 Calculating the commutators of $\tilde Q_3 =\p_{\tilde t}$ with $\tilde Q_1$ and  $\tilde Q_2$ and equating the resulting expressions  to a linear combination of generators (\ref{t3}) (which have to form a basis of the Lie algebra) we obtain a system of two first order linear equations for functions $f(\tilde t)$ and $g(\tilde t)$, whose solutions are  linear, quadratic, exponential  or trigonometric functions of $\tilde t$.  This observation fixes the  possible time dependence of functions $\xi^0$ included into symmetries (\ref{so}).
\section{Equivalence transformations}
\subsection{General speculations}
In addition to symmetries we consider also
 equivalence transformations for equation (\ref{se}) which keep its generic form but can change  exact forms of the arbitrary elements, i.e., of scalar potential $A^0$ and components $A^1, A^2$ of the vector potential. Whenever these elements are not fixed, the  equivalence transformations include a group which can be found in a regular way using the Lie infinitesimal approach. Of course this group includes symmetry transformations as a subgroup. However, for some particular potentials there exist the additional equivalence transformations which have to be fixed also.

To describe all admissible equivalence relations we will use the following fact: the transformed equation should have the same generic form of symmetry operators as the untransformed one, and this form  is given by equations (\ref{so}) and (\ref{kil}).

Let us start with  transformations of the time variable which were discussed in the previous section. They are a constituent part of generic
equivalence transformations which should include also the appropriate changes of the spatial variables $\bf x $  and wave function $\psi(t,{\bf x})$. Since $\xi^0$ in (\ref{so}) should be independent on spatial variables, the admissible changes of  the spatial variable look as follows:
 \begin{gather}\la{t5} t\to \tilde t=F^{-1}(t),\quad t=F(\tilde t).\end{gather}

 To describe functions $F(t)$ we start
 with (\ref{t2}). As it is shown in the previous section
 we should restrict ourselves to such transformations (\ref{t5}) which result in appearance of  new realisation (\ref{t3}) that include either second order polynomial, or  trigonometric, or exponential functions $f(\tilde t), g(\tilde t)$ and $h(\tilde t)$.

 The case of  second order polynomials  by definition  generates operators in  the canonical form (\ref{t2}). The corresponding transformations are generated by the infinitesimal operators (\ref{t2}) and have the following generic form
 \begin{gather}\la{t4} t\to \tilde t=\frac{\nu t +\mu}{t+\lambda}\end{gather}
 where $\mu, \nu$ and $\lambda$ are transformation parameters.

  Transformations (\ref{t4}) can belong to symmetry group of equation (\ref{se}) provided they are accompanied by the appropriate change of the spatial variables and wave function.  On the other hand, whenever $f(\tilde t), g(\tilde t)$ and $h(\tilde t)$ are trigonometric or exponential functions the corresponding transformation (\ref{t5}) is not a symmetry but can  belongs to the equivalence groupoid of (\ref{se}). To obtain the corresponding  transformed generators   we have  three qualitatively different  possibilities for  function $F(\tilde t)$:
 \begin{gather}\la{t6}F=\tan(\omega \tilde t),\  F=\tanh(\omega \tilde t)\end{gather}
 and
 \begin{gather}\la{t7}F(\tilde t)=\text{e}^{2\omega \tilde t}\end{gather}
 with some numeric parameter $\omega$. A more general form of the considered transformation is given by products of (\ref{t6}), (\ref{t7}) and (\ref{t4}).

 Just the changes (\ref{t6}) together with the corresponding changes of independent variable $\bf x$ and dependent variable $\psi(t,\bf x)$ were used in \cite{Nied3} to prove the  equivalence of the free particle and the oscillator SEs.  Notice that the related transformations for the wave function are  rather cumbersome.
 \subsection{Special   transformations and their group-theoretical content}

Let us consider in detail the change of the time variable presented by formula (\ref{t7}). This change keeps the generic form of the equation fixed by formulae (\ref{se}) and (\ref{H2}) provided the dependent and independent variables are transformed in the following manner:
  \begin{gather}\la{t8} t=\text{e}^{2\omega \tilde t},\quad{\bf x}=\sqrt{2\omega}\text{e}^{\omega \tilde t}\tilde{\bf x},
 \quad \psi(t,{\bf x})=\text{e}^{\ri\omega(\tilde{\bf x}^2-\ri t)}\tilde\psi(\tilde t,\tilde{\bf x}).\end{gather}

 Transformations  (\ref{t8}) connect the free particle and repulsive oscillator equations. They can be applied also for more general system restricted by the requirement of the conformal invariance. Notice also that the inverse transforms look as follows:
  \begin{gather}\la{t9}\tilde t=\ln(t)/2\omega,\quad \tilde{\bf x}={\bf x}/{\sqrt{2\omega t}}\end{gather}
  and
  \begin{gather}\la{t99}  \tilde \psi(\tilde t,{\bf \tilde x})=t^{-\frac{3}2}\text{e}^{-\ri{\bf x}^2/t}\psi(t,{\bf x}).\end{gather}

The more cumbersome transformations   proposed in  \cite{Nied3}, are products of   (\ref{t8}) and  transformations (\ref{t4}) with $\nu=\lambda=-\mu=1$.

  Let us remind that the components of vector $\frac{\bf x}{\sqrt{t}}$  are  nothing but  invariants of the dilatation transformations whose generator for the free SE can be written as follows: $ X=2t\p_t+x_a\p_a$ .  And just these invariants change the standard spatial variables in transformations (\ref{t9}).

  Notice that up no normalization  function (\ref{t99}) can be rewritten as
  \begin{gather*} \tilde \psi(\tilde t,{\bf \tilde x})=\psi_f\psi(t,{\bf x}).\end{gather*}
  where $\psi_f=(2\sqrt{2\pi})^{-3}t^{-\frac{3}2}\text{e}^{-\ri{\bf x}^2/t}$ is the fundamental solution of the free SE. Thus our equivalence transformations for the wave function is reduced to its multiplication by  the fundamental solution while the spatial  variables are changed to the invariants of the dilatation transformation.

  Transformations (\ref{t8}) are seemed be nice and have a clear group-theoretical meaning which is discussed in the following paragraphs.

It is pretty well known (see, e.g., \cite{olver}) that by the appropriate change of variables any generator $X$ of a  one parametric Lie group in $ n$-dimensional space
can be reduced to the generator of shifts. Such changed  variables should include  $n-1$ invariants of the group transformations and  variable $\tilde x_n$ satisfying the condition  $X \tilde x_n=1$. As a result $X$ is transformed to $\tilde X=\p_{x_n}$.

Just such procedure is realized by transformations (\ref{t9}) which reduce the dilatation operator  to the form $X=\p_{\tilde t}$. In addition, the corresponding transformations of the wave function are reduced to its expression  via the fundamental solution of the free SE. Thus transformations  (\ref{t9}) do have a nice group theoretical sense.

\subsection{Equivalence transformations which keep  the time variable invariant}
  Consider now admissible transformations which do not change the time variable but affect spatial variables. Their generic form is given by the following formula:
  \begin{gather}\la{tt1}{x }_a\to \tilde  { x}_a={ G}_a({\bf x})+ {R}_a(t).\end{gather}
  It is easy to show by the direct verification that functions   ${R}_a(t)$ are restricted to be  second order polynomials in $t$:
  \begin{gather}\la{tt2}R_a=\mu_a+\nu_a t+\varkappa_a t^2.\end{gather} Parameters $\mu_a$ and $\nu_a$ generate shifts of spatial variables and Galilei transformations respectively and so can be neglected in search for the equivalence relations.   To keep the derivative terms in (\ref{se}) unchanged the related transformation  (\ref{tt1}) with non-trivial $\varkappa$ has to be accompanied  by the following transformations:
   \begin{gather}\la{et3}\begin{split}& x_a\to x'_a=x_a-\frac12{\kappa_a}t^2-\mu,\quad  t\to t'=t,\\& \psi(t,{\bf x})\to\psi'(t',{\bf x}')=\exp\left(-it\kappa_ax_a+\frac\ri3\kappa^2t^3\right)\psi(t,{\bf x})\end{split}\end{gather}
   which connect  the SE with trivial and free fall potentials \cite{Nied3}.

   Let us describe admissible functions ${ G}_a({\bf x})$ in (\ref{tt1}).
   To keep the generic form of symmetries (\ref{so}) and (\ref{kil}) these function   should be linear. To keep the second order derivation terms in SE, the equivalence  transformations should keep the form $x_1^2+x_2^2+x_3^2$ invariant up to multiplication on a constant factor. In other words,  transformations (\ref{tt1}) with trivial $R(t)$ belong to the Euclid group extended by the dilatation transformations. We will denote this group as $\tilde {\textsf{E}}$ and its Lie algebra as $\tilde{\textsf{e}}$(3).

  Thus we describe the equivalence transformations which can be admitted by the classified equations.
Now we are in a position to find admissible symmetries and the
related potentials by integrating the determining equations. It will be done in  the following sections.

\section{Symmetries with second order polynomial dependence on time}
\subsection{Preliminary notes}
Let us start with symmetries whose time derivative terms have the canonical form (\ref{t2}) or are trivial.  Substituting (\ref{kil}) with quadratic, linear, constant and trivial $\xi^0$   into (\ref{so}) we come to the linear combination
\begin{gather}\la{soka} Q=\mu A+\kappa D+\theta^{ab}M_{ab}+\lambda^aG_a \nu^aP_a+\eta\end{gather}
 of symmetries presented in the following formulae:
\begin{gather}\begin{split}& P_a=-\ri\p_a,\quad M_{ab}=\varepsilon_{abc}L_c=x_aP_b-x_bP_a,\\ &D=2tP_0-x_aP_a+{3\ri }/2, \end{split}\label{sok}\\\begin{split}
 &P_0=\ri\p_t,\quad G_a=tP_a-x_a,\\&
A=tD-t^2P_0+{r^2}/2.\end{split}\la{sos}\end{gather}
Here $\varepsilon_{abc}$ is  Levi-Civita symbol,  $\eta$ is some unknown function and  $r^2=x_1^2+x_2^2+x_3^2$.

 A well known system which admits all symmetries
(\ref{sok}) and (\ref{sos}) is the SE with trivial potentials. These symmetries
 together with the unit operator  form a basis of the 13-dimensional Lie algebra $\textsf{schr}(1,3)$ called Schr\"odinger algebra.   Let us present the commutation relations for its basis elements:
 \begin{gather}\la{shr}\begin{split}&[P_a,L_b]=\ri\varepsilon_{abc}P_c, \quad [L_a,L_b]=\ri\varepsilon_{abc}L_c,\\&[P_0,G_a]=\ri P_a,\quad [P_a,G_b]=\ri\delta_{ab}I,\\& [P_a, D]=\ri P_a,\quad [D,G_a]=2\ri G_a,\ [P_a,A]=\ri G_a,\end{split}\\\la{shr1}  [P_0, D]=2\ri P_0,\quad [D,A]=2\ri A,\quad [P_0,A]=\ri D,\end{gather}
 the remaining commutators are equal to zero.
Here  $\delta_{ab}$ is Kronecker delta and $I$  is the unit operator.

  Algebra $\textsf{schr}(1,3)$ includes the subalgebra spanned on symmetries (\ref{sok}). The latter commute with the free Schr\"odinger hamiltonian and are generators of the 3d Euclid group extended by the dilatation transformations.  Whenever the potentials in (\ref{se}) are nontrivial, operators (\ref{sok}) generate transformations which can change them, but keep the generic form of equation (\ref{se}), refer to Section 5.

  Thus operators  (\ref{sok}) generate the   equivalence  group for the mentioned equation.   Nevertheless, for some particular potentials we can find the additional equivalence transformations which do not belong to the equivalence group.

Let us fix some other high dimensional subalgebras of algebra $\textsf{sch}(1,3)$. Deleting basis elements $A$ and $D$ we reduce $\textsf{sch}(1,3)$  to the Lie algebra $\textsf{g}(1,3)$ which generates the Galilei group. One more subalgebra of the same dimension which we denote as $\textsf{A}_{11}$  can be obtained if we delete $P_0$ and $A$.

To find potentials compatible with  symmetries (\ref{sok}) we are supposed to solve the system of equations (\ref{eqA}) and (\ref{con10b}) with {\it constant} parameters $\theta^{ab}=-\theta^{ba}, \nu^a$ and $\alpha$. However, in this case we have as much as seven arbitrary parameters, which make it impossible to integrate (\ref{eqA}) and (\ref{con10b}) directly. Our strategy will be to restrict ourselves to inequivalent reduced combinations of these parameters, and it can be done in a regular way.

Operators (\ref{sok})  form a basis of the extended   Euclid  algebra $\tilde{\text{e}}$(3) whose non-equivalent subalgebras has been classified in \cite{baran}.  And just these subalgebras generate non-equivalent linear combinations of symmetries which we have to consider.

In particular, algebra $\tilde{\text{e}}$(3) has four non-equivalent one-dimensional subalgebras whose  basis elements are: \cite{baran}:
\begin{gather}\la{1d}L_3=M_{12},\quad L_3+P_3,\quad D+\mu L_3,\quad P_3.\end{gather}
The corresponding parameters in (\ref{kil}) are
$\theta^{12}=1$ for $L_3$, $\theta^{12}=\nu^3=1 $ for $L_3+P_3$,
$\alpha=-2, \ \theta^{12}=\mu$ for $D+\mu L_3$ , $\nu^3=1$ for $P_3$.
Substituting these data into  (\ref{con10}) and (\ref{de8}) we come to exactly solvable equations whose integrals give us the list of inequivalent potentials corresponding to the one dimensional subalgebras of $\tilde {\textsf{e}}(3)$.

The next step is to use the non-equivalent two-dimensional subalgebras of $\tilde{\text{e}}$(3). In accordance with  \cite{baran}, it is sufficient to consider the  subalgebras spanned on the following basis elements:
\begin{gather}\la{2d} \langle L_3, P_3 \rangle,\quad \langle D+\kappa L_3, P_3 \rangle,\quad \langle P_2, P_3 \rangle,\quad \langle D, L_3 \rangle.\quad\end{gather}

 Any sets (\ref{2d}) includes at least one  element from (\ref{1d}). Thus we can use the results obtained at the previous step and apply to them the  additional restrictions which are nothing but  equations  (\ref{con10}) and (\ref{de8}) generated by the second basis elements of the considered algebra.

Analogously, considering the non-equivalent three dimensional subalgebras of $\tilde{\text{e}}$(3),whose basis elements are presented in the following formulae
\begin{gather}\la{3dim} \langle D, P_3, L_3\rangle,\quad  \langle D, P_1,  P_2\rangle, \quad \langle L_1, L_2, L_3\rangle,\quad \langle L_3, P_1,  P_2\rangle \\\langle P_1, P_2,  P_3\rangle,\quad \langle L_3+ P_3,\ P_1,\ P_2\rangle,\quad \langle D+\mu L_3, P_1,  P_2\rangle,\ \mu>0\la{empty}\end{gather}
we obtain potentials compatible with three dimensional subalgebras, and so on.
\subsection{Solution of the determining equations}
 Let us start with one dimension subalgebras (\ref{1d}).  For symmetry $Q=P_3+K$ the only nonzero parameter in  (\ref{eqA})   is
$\nu^3$, so the latter equations are reduced   to the following form:
\begin{gather} \la{e1}K_1=-A^1_3,\quad K_2=-A^2_3\end{gather}
and
\begin{gather} \la{e2}K_3=0.\end{gather}

In accordance with  (\ref{e1}) and (\ref{e2}) functions  $A^1$ and $A^2$  are linear in $x_3$, i.e.,
\begin{gather}\la{e3}A^1=x_3a^1+b^1,\quad A^2=x_3a^2+b^2\end{gather}
where $a^1,\ a^2,\ b^1$ and $b^2$ are functions of $x_1$ and $x_2$ of generic form  (\ref{vp}). In particular,
\begin{gather}\la{e01}a^1=F_1+G_2,\quad a^2=F_2-G_1\end{gather}
where $F$ and $G$ are functions of $x_1$ and $ x_2$.

The immediate consequence of (\ref{e1})--  (\ref{e01}) is that function $G$  has to satisfy the condition $G_{11}+G_{22}=0$. Thus there exist a function $Q$ satisfying the Caushy-Riemann condition together with $G$, and functions (\ref{e01}) can be represented as follows:
\begin{gather}\la{e02}a^1=\tilde F_1,\quad a^2=\tilde F_2\end{gather}
where $\tilde F=F+Q$.

Thus the generic form of the vector-potential in equation (\ref{se}) admitting symmetry $Q=P_ 3+K$ is given by formula
(\ref{e3}) where $a^1$ and $a^2$ has the form  (\ref{e02}) while  $b^1$ and $b^2$ are arbitrary functions of  $x_1$ and $x_2$.  Using the gauge transformation these functions can be reduced to $b^1=R_2$ and $b^2=-R_1$ where $R$ is a function of $x_1$ and $x_2$, then
\begin{gather}\la{e8}A^1=\tilde F_1x_3+R_2,\quad A^2=\tilde F_2x_3-R_1,\quad A^3=0.\end{gather}
Moreover, in accordance with (\ref{e1}) the corresponding function $K$ is equal to $-\tilde F$ up to a constant term.

Solving the corresponding equation (\ref{con10b}) where the only non-zero parameter is $\nu^3$ we find that $A^0$ is an arbitrary function of $x_1$ and $x_2$, i.e., $A^0=R(x_1,x_2)$. The above obtained results are presented  in Item 1 of Table 1.

For symmetry $Q=L_3+K$
 we have in (\ref{eqA}) $\alpha=0, \nu^a=0,
 \theta^{13}=\theta^{23}=0$, but parameter $\theta^{12}$ is nonzero. The related equations
 (\ref{eqA}) are reduced to the following form:
 \begin{gather}\la{e03}K_1=A^1_\varphi+A^2,\quad K_2=A^2_\varphi-A^1,\quad K_3=0,\end{gather} while equations  (\ref{con12}) and  (\ref{con13}) look as  follows:
\begin{gather}\la{e6}K_\rho=(\tilde x\cdot A)_\varphi, \quad K_\varphi=(\tilde x\times A)_\varphi \end{gather}
where we use the cylindrical variables $\rho=\ln(\tilde r), \tilde r=\sqrt{x_1^2+x_2^2}, \varphi =\arctan(\frac{x_2}{x_1})$ and denote
\begin{gather}\la{e7}\tilde x\cdot A=x_1A^1+x_2A^2,\quad \tilde x\times A =x_1A^2-x_2A^1.\end{gather}

It follows from (\ref{e03}) that $A^1_{3\varphi\varphi}=-A^1_3$ and $A^2_{3\varphi\varphi}=-A^2_3$, and so the generic form of $A^1$ and $A^2$ looks as follows:
 \begin{gather}\la{e04}\begin{split}&A^1=a^1(\varphi,\rho)+x_1g^1(x_3,\rho)+x_2g^2(x_3,\rho),
 \\&A^2=a^2(\varphi,\rho)+x_2g^1(x_3,\rho)-x_1g^2(x_3,\rho)\end{split}\end{gather}
 where $a^1, a^2, g^1$ and $g^2$ are functions of the arguments fixed in the brackets. Moreover, taking into account the compatibility condition for system (\ref{e03})
  \begin{gather*}(A^1_2-A^2_1)_\varphi=0\end{gather*} and applying the gauge transformation  we can reduce $a^1$,  $a^2$ and functions (\ref{e04}) to the following form: $a^1=x_2G(\rho), a^2=-x_1G(\rho)$  and
  \begin{gather}\la{No1}A^1=x_1R^1(\rho,x_3)+{x_2}R_2(\rho, x_3),\quad A^2=-x_1R^2(\rho,x_3)+x_2R^1(\rho,x_3)\end{gather}
 where $R^1=g^1+G$ and $R^2=g^2+G$ are arbitrary functions
 of  the arguments specified in the brackets. The related function $K$ which have to solve equations  (\ref{e7}) is independent on spatial variables but can depend linearly on $t$.

Symmetry $L_3+\kappa t$ is valid for equation (\ref{se}) with
vector-potential (\ref{No1}) and scalar potential $A^0=R(r) + \kappa \varphi$. The latter can be found by direct integration of equation  (\ref{con10b}) where the only nonzero parameter is $\theta^{12}$.
The obtained potentials are placed in Item 2 of Table 1 and Item 1 of Table 3.

The next symmetry from the list (\ref{1d})  we consider is $Q_3=L^3+P^3$. The related equations  (\ref{eqA}) and  (\ref{con12}), (\ref{con13}) are reduced to the following forms:
 \begin{gather}\la{e9}K_1=A^1_{\tilde\varkappa} +A^2,\quad K_2=A^2_{\tilde\varkappa}-A^1,\quad K_3=0,\end{gather}
 and
 \begin{gather}\la{e10}K_\rho=(\tilde x\cdot A)_{\tilde\varkappa}, \quad K_{\tilde\varkappa}=(\tilde x\times A)_{\tilde\varkappa} \end{gather}
 where ${\tilde\varkappa}=\varphi+x_3$ and notations (\ref{e7}) are used.

 Equations (\ref{e9}) and  (\ref{e10}) are rather similar to  (\ref{e03}) and  (\ref{e6}), but the derivatives w.r.t. $\varphi$ are now changed by the derivatives w.r.t. ${\tilde\varkappa}$. Thus, in complete analogy with the above we obtain the related components of the vector-potential in the following form (compare with (\ref {No1})):
   \begin{gather}\la{No10}A^1=x_1 g^1(\rho,\tilde \kappa)+{x_2} g_2(\rho,\tilde \kappa),\quad A^2=-x_1 g^2(\rho,\tilde \kappa)+x_2 g^1(\rho,\tilde \kappa) \quad  A^3=0\end{gather}
   where $g^1$ and $g^2$ are arbitrary functions of $\rho=\ln(\tilde r)$ and $\kappa=\arctan(\frac{x_2}{x_1})-x_3$. The related function $K$ is again independent on spatial variables.

 The last symmetry from the list (\ref{1d})  which we have to consider  is $Q_4=D+\mu L^3$. The related equations  (\ref{eqA}) are:
 \begin{gather}\la{e12}K_1=A^1+\mu (A^2+2A^1_y),\quad K_2=A^2-\mu (A^1+2A^2_y),\\ \la{e13} K_3=0\end{gather} where we set $a=2$ and
 $y=\mu \ln(\tilde r)+ \varphi.$

 The compatibility condition for system (\ref{e12}) is $K_{12}=K_{21}$, or
 \begin{gather}A^1_2-A^2_1+\mu(A^1_2-A^2_1)_y=0.\la{e14}\end{gather}
 Then, substituting (\ref{vp})  into (\ref{e14}) we obtain the following equation for function $G$:
 \begin{gather}\la{e15}G_{11}+G_{22}+\mu(G_{11}+G_{22})_y=0\end{gather}
 whose generic solution is:
 \begin{gather}\la{e16}G=\Phi\left(\tilde r/r,z\right) + R,\quad z=\mu\rho-\varphi\end{gather}  where $R$ is a solution of the 2d Laplace equation which can be reduced to zero by the gauge transformation.

 In view of (\ref{vp}) and (\ref{e16}) equations  (\ref{e12}) are reduced to the following ones:
 \begin{gather*}K_1=F_{1y},\quad K_2=F_{2y}\end{gather*} and so
 \begin{gather}\la{e17}K=F_y+\phi(x_3).\end{gather}

 Combining (\ref{e17}) with (\ref{e13}) we find functions $F$ and $K$ in the following form:
 \begin{gather*}F=\tilde F(\theta,z)+F(z,y), \quad K=2\mu\p_y F(z,y)\end{gather*}
 which, together with (\ref{vp}) and (\ref{e16}) defines the generic form of the vector-potential admitting symmetry $Q_4$.

In analogous manner we obtain potentials which correspond to more extended symmetries.
 To this effect it is sufficient to start with the solutions found in the above and apply the additional restrictions generated by symmetries which extend the ones  given in (\ref{1d})  to two, three, and higher dimensional algebras. For example, starting with potentials (\ref{No1}) and asking for the additional symmetry $Q_1=P^3$ we obtain the following version of equations (\ref{e1}) and (\ref{e2}):
\begin{gather*}K_1=-x_1R^1_3-x_2R^2_3,\quad K_2=-x_2R^1_3+x_1R^2_3,\\K_3=0\end{gather*}
whose solution is $R^1=x_3F(\rho),\ R^2=\Phi(\rho), \ K=-F(\rho)$.
 In this way we come to vector-potentials  presented in Item 3 of Table 3 where the additional symmetry $G_3$ is indicated also.

The complete
list of such obtained  vector-potentials together with the related symmetries
is given in Tables  1 and 2, where  $G(.,.), F(.,.), \tilde F(.,.)$ are arbitrary  functions of the arguments fixed in the brackets, $\varphi$ and $\theta$ are Euler angles, $ r=\sqrt{x_1^2+x_2^2+x_3^2}, \tilde r=\sqrt{x_1^2+x_2^2}$, $P_a, $, $L_a $, $G_a (a=1,2,3), D$  and $A$ are symmetries listed in (\ref{sok}) and (\ref{sos}).

All systems presented in  Tables 1--4 admit symmetries $P_0=\ri \p_0$ and $I$, the latter is  the unit operator which generates the scaling of the wave function. The additional symmetries are fixed in the third column of the tables. The related symmetry algebras
are specified in the fourth columns, where
 $\textsf{n}_{a,b}$ and $\textsf{s}_{a,b}$ are nilpotent and solvable Lie
 algebras correspondingly, whose  dimension is $a$ and  identification number is $b$.   To identify these algebras for $a\leq 6$ we use the notations proposed in \cite{snob}. The symbol $2\textsf{n}_{1,1}$ denotes the direct sum of two  one-dimension algebras. In addition, $\textsf{g(1,2)}$ and $\textsf{shcr(1,2)}$ are the Lie algebras of  Galilei and  Schr\"odinger groups in (1+2) dimensional space.
\newpage\begin{center}Table 1.
Symmetries   which are second order polynomials in time variable.
\end{center}
\begin{tabular}{l l l l }
\hline No&Vector potentials $V$&Symmetries&Algebras
\vspace{1mm}

\\

\hline
\vspace{2mm}

1$\hspace{0mm} $&$\begin{array}{l}A^1=x_3\p_1F({ x_1, x_2})+ \p_2G({ x_1, x_2}),\\ A^2= x_3 \p_2F({ x_1, x_2})- \p_1G({x_1,x_2}),\\A^0=R(x_1,x_2)\end{array}$&
$P_3-F({x_1,x_2})$& $\hspace{-2mm}3\textsf{n}_{1,1}$\\

\vspace{1mm}

2$\hspace{0mm} $&
$\begin{array}{l}A^1=x_1G^1(\tilde r,x_3)+{x_2}G^2(\tilde r,x_3), \\ A^2=x_2G^1(\tilde r,x_3)-x_1G^2(\tilde r, x_3),\\A^0=R(\tilde r,x_3)\end{array}$&
$L_3$&$\hspace{-2mm}3\textsf{n}_{1,1}$\\

3$\hspace{0mm} $& \vspace{1mm} $\begin{array}{l}A^1=x_1 G^1(\tilde r, \varkappa)+{x_2} G_2(\tilde r, \varkappa),\\ A^2=-x_1G^2(\tilde r, \varkappa)+x_2 G^1(\tilde r, \varkappa),\\A^0=R(\tilde r,\varkappa), \varkappa=\varphi-x_3\end{array}$&$L_3+P_3$&$\hspace{-3.5mm}\begin{array}{c}

3\textsf{n}_{1,1}\end{array}$\\

\vspace{1mm}

4$\hspace{0mm} $&$\begin{array}{l}A^1=x_3\p_1F({\tilde r})+ \p_2G({\tilde r}),\\ A^2= x_3 \p_2F({\tilde r})- \p_1G({\tilde r}),\\A^0=R(\tilde r)\end{array}$&
$P_3-F(\tilde r),\  L_3$& $\hspace{-4mm}\begin{array}{l}4\textsf{n}_{1,1} \end{array}$\\

5*\vspace{1mm}$\hspace{0mm} $&$\begin{array}{l}A^1=x_3 \p_1\left({F(\varphi)}/{\tilde r}\right) +\p_2G(\varphi),\\ A^2=x_3 \p_2\left({F(\varphi)}/{\tilde r}\right) -\p_1G(\varphi),\\A^0=R(\varphi)/{\tilde r^2}\end{array}$&
 $D,
P_3-{F(\varphi)}/{\tilde r}$ &$\hspace{-3mm}\begin{array}{l}\textsf{s}_{3,1}
\oplus\textsf{n}_{1,1}\end{array}$\\

\vspace{1mm}

6$^\star\hspace{0mm} $&$\begin{array}{l}A^1= \p_2G({ x_1, x_2}),\
A^2=- \p_1G({x_1,x_2}),\\A^0=R(x_1,x_2)\end{array}$&
$P_3, G_3 $
& \hspace{-3mm}$\begin{array}{l}\textsf{n}_{4,1}\end{array}$\\

\vspace{1mm}

7*$\hspace{0mm} $
&  \vspace{1mm}$ \begin{array}{l}A^1=x_2F(\theta,\varphi)/r^2+\p_2G(\theta),\\ A^2=-x_1F(\theta ,\varphi)/r^2-\p_1G(\theta),\\A^0=\frac1{ r^2}R(\theta,\varphi)\end{array}$
 \vspace{0mm}
 &$\begin{array}{l}D, \ A \end{array}$&$\hspace{-3mm}\begin{array}{l} \textsf{sl}(2,R)\oplus\textsf{n}_{1,1}\end{array}$\\

 \vspace{2mm}

 8$^\star $&$\begin{array}{l}A^1=F(x_3),\ A^2=G(x_3),\
A^0=R(x_3)\end{array}$&$ \begin{array}{l}P_1,\  P_2
\end{array}$&$\hspace{-2mm}4\textsf{n}_{1,1}$\\

 \vspace{1mm}

9$^\star\hspace{0mm}$\vspace{1mm}&$\begin{array}{l}A^1=F(x_3),\ A^2=0,\ A^0=R(x_3)\end{array}$&$\begin{array}{l} P_1,\ P_2,\ G_2 \end{array}$&$\hspace{-3mm}\begin{array}{l}\textsf{n}_{4,1}\oplus\textsf{n}_{1,1}\end{array}$\\

 \vspace{1mm}
10$^\star$&$\begin{array}{l}A^1=\p_2G({\tilde r}), A^2=- \p_1G({\tilde r}), A^0=R(\tilde r)\end{array}$&
$\begin{array}{l}P_3,  L_3,   G_3 \end{array}$& $\hspace{-4mm}\begin{array}{l} \textsf{n}_{4,1}\oplus\textsf{n}_{1,1}\end{array}$\\

\vspace{1mm}
11*$\hspace{0mm} $
&  \vspace{1mm}$ \begin{array}{l}A^1=x_2F(\theta)/r^2+\p_2G(\theta),\\ A^2=-x_1F(\theta )/r^2-\p_1G(\theta),\\A^0=\frac1{ r^2}R(\theta,\varphi)\end{array}$
 \vspace{0mm}
 &$\begin{array}{l}D, \ A,\ L_3 \end{array}$&$\hspace{-3mm}\begin{array}{l} \textsf{sl}(2,R)\oplus2\textsf{n}_{1,1}\end{array}$\\

\vspace{1mm}
 12*$^\star\hspace{0mm} $&$\begin{array}{l}A^1=\p_2G\left(\tilde r^\mu\exp(-\varphi)\right),\\ A^2=-\p_1G\left(\tilde r^\mu\exp(-\varphi)\right),\\A^0=\frac1{\tilde
r^{2}}R(\tilde r^\kappa
e^{-\varphi})\end{array}$&
 $\begin{array}{l}D+\mu L_3,\\
P_3,\ G_3\end{array}$ &$\hspace{-2mm}\textsf{s}_{5,38}$\\

\vspace{1mm}

13$\hspace{0mm} $&
\vspace{1mm}
$\begin{array}{l}A^1=A^2=0, \ A^0=G(r)\end{array}$&$\begin{array}{l}L_1,\  L_2,\  L_3\end{array}$ &$\hspace{-2mm}\textsf{so}(3)\oplus  2\textsf{n}_{1,1}$\\

\vspace{1mm}

14*$^\star\hspace{0mm} $&$\begin{array}{l}A^1=0,\  A^2= 0,\ A^0=\frac1{\tilde r^2}R(\varphi)\end{array}$&
 $\begin{array}{l}D,\ A,\
P_3,\  G_3 \end{array}$ &$\hspace{-3mm}\begin{array}{l}
\textsf{s}_{6,242}\end{array}$\\

\vspace{1mm}

15*$^\star\hspace{0mm}$\vspace{1mm}&$\begin{array}{l}A^1=\frac{\lambda}{x_3},\ A^2=0,\ A^3= \frac{\mu}{x_3^2}\end{array}$&$\begin{array}{l} P_1,\ P_2,\ G_2 ,\ D \end{array}$&$\hspace{-3mm}\begin{array}{l}\textsf{s}_{6,96}\end{array}$\\

\vspace{1mm}

16*$ $&\vspace{1mm}$\begin{array}{l}A^1=A^2=0, \ A^0=\frac\kappa {r^{2}}\end{array}$& $\begin{array}{l}A,\ D,\\
L_1, \ L_2, \ L_3\end{array}
$&$\hspace{-2mm}\textsf{sl}(2,R)\oplus\textsf{so}(3)
\oplus\textsf{n}_{1,1}$\\

17$^\star\hspace{0mm}$\vspace{1mm}&$\begin{array}{l}A^1=0,\ A^2=0,\ A^0=R(x_3)\end{array}$&$\begin{array}{l} P_1,\ P_2,\ G_1,\\ G_2, \ L_3 \end{array}$&$\hspace{-3mm}\begin{array}{l}\textsf{g}(1,2)\end{array}$\\

\vspace{1mm}

 18*$^\star $&$\begin{array}{l}A^1=0,\ A^2=0,\
A^0=\frac\kappa{x_2^2}\end{array}$ \vspace{1mm} &$
\begin{array}{l}A, \ D,\ P_1,\  P_3, \\G_1,\ G_3,\ L_2\end{array}$&$\hspace{-2mm}
\textsf{schr}(1,2)$\\

\hline\hline
\end{tabular}
\vspace{2mm}

The systems presented in Items 5, 7,  11, 14, 15, 16, 18 (and 12 for $\mu=0$) of Table 1  admit equivalence transformations   (\ref{t8}) which generate the additional potential term $-\frac{\omega^2 r^2}2$ in the scalar potential. To fix the systems admitting such equivalence transformation we use the mark * for the item numbers. In addition, the systems marked by stars admit equivalence transformations  (\ref{et3}) with respect to variables $x_a$ missing in all potentials.

These rules pay also for the following Tables 2-4. However, for the systems presented in Items 3, 4, 7 of Table 2 and Item 6 of Table 4 transformations  (\ref{et3}) should be combined with the gauge transformation deleting $\alpha x_1$ or $\alpha x_2$.

\begin{center}Table 2.
Symmetries   induced by external magnetic fields
\end{center}
\begin{tabular}{l l l l }
\hline No&Vector potentials $V$&Symmetries&Algebras
\vspace{1mm}\\
\hline

\vspace{1mm}

1$\hspace{0mm}\vspace{1mm}
 $
&  $ \begin{array}{l} A^1=\p_1(\tilde F(\theta,z)+F(y,z )\\ +\p_2G(\theta,z),\\A^2=\p_2(\tilde F(\theta,z)+F(y,z) )\\-\p_1G(\theta,{z})\\A^0=\frac1{r^2} R(\theta,{z}),\\ y=\mu\ln(\tilde r)+\varphi,\ z=\mu\ln(\tilde r)-\varphi,
\end{array}$
 \vspace{1mm}
 &$\begin{array}{l}D+\mu L_3+2\mu\p_y F(y,z) \end{array}$&$\hspace{-1mm}\textsf{s}_{2,1}\oplus\textsf{n}_{1,1}$\\

 2$\hspace{0mm}\vspace{1mm}
 $
&  $ \begin{array}{l} A^1=\p_1F(y,z )+\p_2G(\theta,z),\\A^2=\p_2F(y,z) -\p_1G(\theta,{z}),\\A^0=\frac1{r^2} R(\theta,{z}),
\end{array}$
 \vspace{1mm}
 &$\begin{array}{l}D+\mu L_3+2\mu\p_y F(y,z),\\P_3,\ G_3  \end{array}$&$\hspace{-1mm}\textsf{s}_{5,38}$\\

3$^\star$\vspace{1mm}&$\begin{array}{l}A^1=F(x_3)-\alpha x_2,\\ A^2=G(x_3)+\alpha x_1,\\A^0=R(x_3),\ \alpha\neq0\end{array}$&$\begin{array}{l} P_1-\alpha x_2, P_2+\alpha x_1 \end{array}$&$\hspace{-3mm}\begin{array}{l}\textsf{s}_{4,7}\end{array}$\\

4$^\star$\vspace{1mm}&$\begin{array}{l}A^1=-\alpha x_2,\\ A^2=\alpha x_1,\\A^0=R(x_3),\ \alpha\neq0\end{array}$&$\begin{array}{l} P_1-\alpha x_2, P_2+\alpha x_1,\\L_3 \end{array}$&$\hspace{-3mm}\begin{array}{l}\textsf{s}_{4,7}
\oplus\textsf{n}_{1,1}\end{array}$\\

5\vspace{1mm}&$\begin{array}{l}A^1=-\alpha x_2+\nu\cos(x_3),\\  A^2=\alpha x_1+\nu\sin(x_3),\\ A^0=R(\varphi-x_3)\end{array}$
&$\begin{array}{l} P_1-\alpha x_2, P_2+\alpha x_1,\\ L_3+ P_3 \end{array}$&$\hspace{-3mm}\begin{array}{l}\textsf{s}_{4,7}\oplus\textsf{n}_{1,1}\end{array}$\\

6\vspace{1mm}&$\begin{array}{l}A^1={\lambda x_3}\p_1\varphi/r,\ A^2=\lambda x_3\p_2\varphi/r,\\ A^0=R(r)\end{array}$&$\begin{array}{l}L_1+x_3A^2-\lambda x_1/r,\\L_2-x_3A^1-\lambda  x_2/r,\\ L_3\end{array}$&$\hspace{-3mm}\begin{array}{l}\textsf{so}(3)
\oplus2\textsf{n}_{1,1}
 \end{array}$\\

\vspace{2mm}

7\vspace{1mm}&$\begin{array}{l}A^1={\lambda x_3}\p_1\varphi/r,\ A^2=\lambda x_3\p_2\varphi/r,\\ A^0=\mu/{r^2}\end{array}$&$\begin{array}{l}L_1+x_3A^2-\lambda x_1/r,\\L_2-x_3A^1-\lambda  x_2/r,\ \\L_3, D, A\end{array}$&$\hspace{-3mm}\begin{array}{l}\textsf{so}(3)\oplus\textsf{sl}(2,R)\\\oplus
\textsf{n}_{1,1} \end{array}$\\

8$^\star$\vspace{1mm}&$\begin{array}{l}A^1=-\alpha x_2,\  A^2=\alpha x_1,\\ A^0=0\end{array}$
&$\begin{array}{l} P_1-\alpha x_2, P_2+\alpha x_1,\\P_3, G_3, L_3  \end{array}$&$\hspace{-3mm}\begin{array}{l}\textsf{s}_{7,1}\end{array}$\\

\hline\hline
\end{tabular}
\vspace{2mm}

In Item8 of Table 2 a  seven dimensional invariance algebra   is presented. It is solvable, and we conventionally
denote  it as $\textsf{s}_{7,1}$. The commutation relations for its basis elements can be found in (\ref{shr}) and the following formulae (\ref{ttt}).

\section{Symmetries with exponential dependence on time}
 The next step is to solve the determining equations for symmetries with the exponential  dependence on time. The generic form of such symmetries can be obtained starting with (\ref{soka}), (\ref{sok}) and (\ref{sos}) and making transformations (\ref{t7}), (\ref{t8}). In this way we find that up to constant multipliers the following transitions take place:
   \begin{gather}\la{eq1}\begin{split}&P_0\to A^-({\omega}),\quad D\to P_0, \quad A\to A^+({\omega}),\\&P_a\to B^-_a({\omega}), \quad G_a\to B^+_a(\omega),\quad L_a\to L_a\end{split}\end{gather}
   where
   \begin{gather}\la{sos3}\begin{split}& A^{\pm}(\omega)=\exp(\pm2\omega t)(2P_0+\omega^2r^2\mp
{\omega}(x_aP_a+P_ax_a)), \\&B^{\pm}_a(\omega)=
\exp(\pm \omega t)(P_a\mp\omega x_a). \end{split}\end{gather}

The mapping  (\ref{eq1}) is nothing but a
transformation of basis elements of algebra $\textsf{schr}(1,3)$ to the new equivalent forms. However in general this  transformation does not keep the generic form of equation (\ref{se}), and so the description of potentials compatible with symmetries (\ref{sos3}) is a separate part of our classification problem.

Like in the previous section to solve the related equations  (\ref{eqA}) and (\ref{con10b}) we can use the inequivalent subalgebras of the algebra spanned on basis  (\ref{eq1}). However,
since equation (\ref{se}) admits symmetry $P_0=\ri\p_t,$  we have to restrict ourselves to such  linear combinations of  basis elements (\ref{eq1}) which have the same dependence on time. Thus effectively we have to consider the  inequivalent one dimensional subalgebras spanned on the following basis elements:
\begin{gather}\la{e22}B^+_3(\omega)+\exp(\omega t)K,\quad A^+(\omega)+\exp(2\omega t)\tilde K\end{gather}
where we reserve the possible change of the non-derivation part of symmetries caused by the presence of the magnetic field by adding new arbitrary functions $K({\bf x})$ and $\tilde K({\bf x})$.

At this stage we ignore operators $B^-_3(\omega)$ and $A^-(\omega)$ since they coincide with $B^+_3(\omega)$  and $A^+(\omega)$ up to the sign of $\omega$.

Let us search for potentials compatible with the first symmetry from the list (\ref{e22}), i.e., $B^+_3(\omega)+\exp(\omega t)K(t,\bf x).$  The corresponding equations (\ref{eqA}) and (\ref{con10b}) include the only nonzero multiplier $\nu^3=\exp(\omega t)$ for the derivative terms and are reduced to the forms:
\begin{gather}\la{eq3} \p_1K=-\p_3A^1, \quad \p_2K=-\p_3A^2,\quad \p_3K=0\end{gather}
and
\begin{gather}\la{eq4}\p_3A^0=-\omega^2x_3+\omega K.\end{gather}

Equations (\ref{eq3}) are similar to system  (\ref{e1}), (\ref{e2}) whose solutions are given by formulae (\ref{e8}). Substituting these data into equation (\ref{eq4}) and integrating we obtain the corresponding scalar potential:
\begin{gather}\la{eq5} A^0= G({x_1,x_2})-\frac{\omega^2x_3^2}2+\omega x_3 F(x_1,x_2).\end{gather}
These results are presented in Item 1 of Table 4.

\newpage

\begin{center}Table 3.
Symmetries induced by external electric fields
\end{center}
\begin{tabular}{l l l l }
\hline

\vspace{2mm}

 No&Potentials &Symmetries&Algebras
\\

\hline

\vspace{2mm}

1$\hspace{0mm} $& \vspace{1mm}
$\begin{array}{l}A^1=x_1G^1(\tilde r,x_3)+{x_2}G^2(\tilde r,x_3), \\ A^2=x_2G^1(\tilde r,x_3)-x_1G^2(\tilde r, x_3),\\A^0=R(\tilde r,x_3)+\kappa\varphi\end{array}$&
$\begin{array}{l}L_3+\kappa t \end{array}$&\hspace{-3.5mm}$\begin{array}{c}\textsf{n}_{3,1}\end{array}$\\

2$\hspace{0mm} $& \vspace{1mm} $\begin{array}{l}A^1=x_1 G^1(\tilde r, \varkappa)+{x_2} G_2(\tilde r, \varkappa),\\ A^2=-x_1G^2(\tilde r, \varkappa)+x_2 G^1(\tilde r, \varkappa),\\A^0=R(\tilde r,\varkappa)+\kappa \varphi, \varkappa=\varphi-x_3\end{array}$&$\begin{array}{l}L_3+P_3+\kappa t\end{array}$&$\hspace{-3.5mm}\begin{array}{c}\textsf{n}_{3,1}\end{array}$\\

3$^\star$&$\begin{array}{l}A^1=\p_2G({\tilde r}), A^2=- \p_1G({\tilde r}),\\A^0=R(\tilde r)+\kappa\varphi\end{array}$&
$\begin{array}{l}P_3,\  L_3+\kappa t,  \ G_3 \end{array}$& $\hspace{-4mm}\begin{array}{l} \textsf{s}_{5,14} \end{array}$\\

4$\hspace{0mm} $&$\begin{array}{l}A^1=x_3\p_1F({\tilde r})+ \p_2G({\tilde r}),\\ A^2= x_3 \p_2F({\tilde r})- \p_1G({\tilde r}),\\A^0=R(\tilde r)+\kappa\varphi\end{array}$&
$\begin{array}{l}P_3-F(\tilde r),\\  L_3+\kappa t\end{array}$& $\hspace{-4mm}\begin{array}{l}\textsf{n}_{3,1}\oplus\textsf{n}_{1,1}\end{array}$\\

5&$\vspace{0mm}\begin{array}{l}A^1= \p_2G({ x_1, x_2}),\\ A^2= - \p_1G({x_1,x_2})\\A^0= R({x_1,x_2})-\frac{\omega^2x_3^2}2\end{array}$&$\begin{array}{l}B^+_3(\omega) ,\ B^-_3(\omega) \end{array}
$&$\hspace{-3mm} \begin{array}{l} \textsf{s}_{4,6} \end{array}$\\

6\hspace{2mm}
 \vspace{2mm}
 &$\begin{array}{l}A^1= \p_2G({\tilde r}),\ \ A^2= - \p_1G({\tilde r})\\A^0=R({\tilde r})-\frac{\omega^2x_3^2}2\end{array}$&
$\begin{array}{l}B^+_3(\omega),\  B^-_3(\omega)
,\\ L_3 \end{array} $&$\hspace{-3mm} \begin{array}{l}\textsf{s}_{4,6}\oplus\textsf{n}_{1,1} \end{array}$\\

7$^\star$\hspace{1mm} \vspace{1mm}&$\begin{array}{l}A^1=0, \  A^2=G(x_1) ,\\A^0=R({x_1})-\frac{\omega^2x_3^2}2\end{array}$&
$\begin{array}{l}B^+_3(\omega), \ B^-_3(\omega),\\ P_2
\end{array} $&$\hspace{-3mm} \begin{array}{l}\textsf{s}_{4,6}\oplus\textsf{n}_{1,1} \end{array}$\\

9\hspace{1mm} \vspace{1mm}&$\begin{array}{l}A^1=0,\ A^2=0,\\A^0=R(x_3)-\frac{\omega^2\tilde r^2}2\end{array}$&$\begin{array}{l} B^+_1(\omega_1),\  B^-_1(\omega_1),\\ B^+_2(\omega_2),\ B^-_2(\omega_2),\  L_3 \end{array}$&$\hspace{-3mm} \begin{array}{l}\textsf{s}_{7,2} \end{array}$\\

10$^\star$\hspace{1mm} \vspace{1mm}&$\begin{array}{l}A^1=A^2=0,\\ A^0=-\frac{\omega_1^2x_1^2}2 -\frac{\omega_2^2x_2^2}2 \end{array}$&$\begin{array}{l}  B^+_1(\omega_1),\  B^-_1(\omega_1),\\ B^+_2(\omega_2),\ B^-_2(\omega_2) , \\  P_3, \ G_3 \end{array}$&$\hspace{-3mm} \begin{array}{l}\textsf{s}_{8,1} \end{array}$\\

 11\vspace{1mm}&$\begin{array}{l}A^1= A^2=0,\\A^0=-\frac{\omega_1^2x_1^2}2-
\frac{\omega_2^2x_2^2}2-\frac{\omega_3^2x_3^2}2\end{array}$
&$\begin{array}{l} B^+_1(\omega_1),\  B^-_1(\omega_1),\\ B^+_2(\omega_2),\ B^-_2(\omega_2) ,\\ B^+_3(\omega_3),\ B^-_3(\omega_3) \end{array}$&$\hspace{-3mm} \begin{array}{l}\textsf{s}_{8,2} \\ \end{array}$\\

12$^\star$\hspace{1mm} \vspace{1mm}&$\begin{array}{l}A^1=  A^2=0, \\ A^0=-\frac{\omega^2x_3^2}2\end{array}$
&$\begin{array}{l} P_1\  P_2,\ G_1,\ G_2\\ B^+_3(\omega), \ B^-_3(\omega), \ L_3 \end{array}$&$\textsf{s}_{9,1} $\\

 13$^\star$\hspace{1mm} \vspace{1mm}&$\begin{array}{l}A^1=A^2=0,\\ A^0=-\frac{\omega^2\tilde r^2}2\end{array}$&$\begin{array}{l}  B^+_1(\omega_1),\  B^-_1(\omega_1),\\ B^+_2(\omega_2),\ B^-_2(\omega_2) , \\L_3, \ P_3, \ G_3 \end{array}$&$\hspace{-3mm} \begin{array}{l}\textsf{s}_{9,2} \end{array}$\\

14\vspace{1mm}&$\begin{array}{l}A^1= A^2=0,\\A^0=-\frac{\omega^2\tilde r^2}2-
\frac{\omega_3^2x_3^2}2\end{array}$
&$\begin{array}{l} B^+_1(\omega_1),\  B^-_1(\omega_1),\\ B^+_2(\omega_2),\ B^-_2(\omega_2) ,\\ B^+_3(\omega_3),\ B^-_3(\omega_3) ,\ L_3 \end{array}$&$\hspace{-3mm} \begin{array}{l}\textsf{s}_{9,3} \\ \end{array}$\\
\hline\hline
\end{tabular}

\vspace{4mm}

For symmetry $A^+(\omega)+\exp(2\omega t)\tilde K$ we have the following form of the determining equations  (\ref{eqA}) and (\ref{con10b}):
\begin{gather}\la{eq6}\p_1K=\omega(\p_\rho+x_3\p_3+1)A^1,\quad \p_2K=\omega(\p_\rho+x_3\p_3+1)A^2,\quad \p_3 K=0\end{gather}
and
\begin{gather}\la{eq7}-(\p_\rho+x_3\p_3)A^0+2\omega(x_1A^1+x_2A^2)=2 A^0+\omega^2(\exp(2\rho)+x_3^2)+2\omega  K\end{gather}
where we use the cylindrical variables $\rho=\ln(x_1^2+x_2^2), \varphi =\arctan\left(\frac{x_2}{x_1}\right)$.

Equations (\ref{eq6}) coincide with the system (\ref{e12}), (\ref{e13}) with trivial parameter $\mu$. Thus we have the following solutions for $A^1, A^2$ and $K$:
\begin{gather}\la{eq8}\begin{split}& A^1=\p_1 F(\theta,\varphi)+\p_2G(\theta,\varphi),\\&  A^2=\p_2 F(\theta,\varphi)-\p_1G(\theta,\varphi),\quad K=0\end{split}\end{gather}

Substituting (\ref{eq8}) into (\ref{eq7}) we reduce the latter equation to the following form:
\begin{gather}\la{eq9}-(\p_\rho+x_3\p_3)A^0=2 A^0+\omega^2(\exp(2\rho)+x_3^2)-\omega(2\p_\varphi G(\theta,\varphi)-\sin(2\theta)\p_\theta F(\theta,\varphi))\end{gather}
and so
\begin{gather}\la{eq11}A^0= -\frac{\omega^2 r^2}2+\frac{R(\theta,\varphi)}{r^2}+\omega(\sin(2\theta)\p_\theta F(\theta,\varphi)/2-\p_\varphi G(\theta,\varphi)).\end{gather}
 These results are presented in Item 10 of Table 4.

Notice that functions $F(.,.)$  in Tables 2, 4 and parameters $\kappa$ in Tables 3, 4 are supposed to be nontrivial.

Thus we find equations (\ref{se}) invariant w.r.t. one dimensional algebras spanned on basis elements (\ref{e22}). The corresponding potentials include arbitrary functions. For some particular versions of these functions the equation under study admits more extended symmetry algebras. We will not present here the requested  routine calculations which, in analogy with the previous section, can be made going over inequivalent subalgebras of algebra $\tilde{e}(3)$. The obtained results are summarized in Tables 3 and 4.

The dimensions of symmetry algebras presented in Items 8--14 of Table 3 is higher than 6. The classification of such algebras is still missing, it is a "wild" algebraic problem. Thus  all we can do here is to fix their dimension, solvability and  present explicitly the nonzero  commutation relations for  basis elements.  The latter ones can be found in the following formulae:
\begin{gather}\la{ttt}\begin{split}&
[B_a^-(\omega),B_a^+(\omega)]=2\ri\omega I,\quad [\hat P_2, \hat P_1]=2\ri\alpha I, \\&[P_0, B^{\varepsilon}_a(\omega)]=\ri \varepsilon \omega I,\
[L_3,\hat P_1]=\ri \hat P_2, \quad [L_3,\hat P_2]=-\ri \hat P_1,\\ &
[L_3,B^{\varepsilon}_1(\omega)]=\ri B^{\varepsilon}_2(\omega), \quad
[L_3,B^{\varepsilon}_2(\omega)]=-\ri B^{\varepsilon}_1(\omega)
\end{split} \end{gather}
where $ \hat P_1=P_1-\alpha x_2$  and $ \hat P_2+\alpha x_1$.  The remaining commutation relations including also $P_a$ and $G_a$ can be found in (\ref{shr}).

In accordance with (\ref{ttt}) algebras $\textsf{s}_{m,n}$ with dimension $m>6$ indicated in Table 3 include  two or three Heisenberg  subalgebras $\textsf{n}_{3,1}$ which are ideals.

 The next and the last table includes systems whose symmetries are induced by superpositions of electric and magnetic fields. The dimensions of the related symmetry algebras do not exceed six. Any of the systems presented in Table 4 admits at least one symmetry exponential in the time variable. The commutation relations for symmetries can be found in (\ref{ttt}), (\ref{shr}) and the following formulae:
\begin{gather*}\begin{split}&[P_0,A^+(\omega)]=2\ri\omega A^+(\omega),\\&[A^+(\omega),B^+_3+K]=0,\\& [A^+(\omega),B^-_a]=2\ri B^+_a\end{split}\end{gather*}
where  $K$ denotes function added to $B^+_3$ in Items 12-14. In particular, in Item 14 we have $K=\frac{\text{e}^{\omega t}}{ x_1}$.
\newpage

\begin{center}Table 4.
Symmetries induced by superpositions of electric and magnetic external fields
\end{center}
\begin{tabular}{l l l l }
\hline

\vspace{1mm}

 No&Potentials $A^a$&Symmetries&Algebras
\\
\hline

\vspace{-3mm}&&&\\
1&\vspace{2mm}$\begin{array}{l}A^1=x_3\p_1F({ x_1, x_2})+ \p_2G({ x_1, x_2}),\\ A^2= x_3 \p_2F({ x_1, x_2})- \p_1G({x_1,x_2})\\A^0= G({x_1,x_2})-\frac{\omega^2x_3^2}2+\omega x_3 F(x_1,x_2)\end{array}$&$\begin{array}{l}B^+_3(\omega)-\text{e}^{\omega t} F(x_1,x_2) \end{array}
$&$\hspace{-3mm} \begin{array}{l}\textsf{s}_{2,1}\oplus\textsf{n}_{1,1}\end{array}$\\
\vspace{2mm}

2\hspace{2mm}
 \vspace{0mm}
 &$\begin{array}{l}A^1= \p_2G({\tilde r}),\ \ A^2= - \p_1G({\tilde r})\\A^0=R({\tilde r})-\frac{\omega^2x_3^2}2+\kappa \varphi\end{array}$&
$\begin{array}{l}B^+_3(\omega),\  B^-_3(\omega)
,\\ L_3+\kappa t \end{array} $&$\hspace{-3mm} \begin{array}{l}\textsf{s}_{5,15}\end{array}$\\

3
\hspace{1mm}\vspace{0mm}

 &$\begin{array}{l}A^1=x_3\p_1F({ \tilde r})+ \p_2G({\tilde r}),\\ A^2= x_3 \p_2F({ \tilde r})- \p_1G({\tilde r})\\A^0=R({\tilde r})-\frac{\omega^2x_3^2}2+\omega x_3 F(\tilde r)+\kappa \varphi\end{array}$&
$\begin{array}{l}B^+_3(\omega)-\text{e}^{\omega t}F(\tilde r)
,\\ L_3+\kappa t \end{array} $&$\hspace{-3mm} \begin{array}{l}\textsf{s}_{2,1}\oplus2\textsf{n}_{1,1}\end{array}$\\

\vspace{2mm}

4$^\star$\hspace{2mm} \vspace{0mm}&$\begin{array}{l}A^1=x_3\p_1F({ x_1}),\  A^2= G({x_1}),\\A^0=R({x_1})-\frac{\omega^2x_3^2}2+\omega x_3 F(x_1)\end{array}$&
$\begin{array}{l}B^+_3(\omega)-\text{e}^{\omega t} F(x_1), \\ P_2,\text{ and } G_2\text{ if }A_2=0
\end{array} $&$\hspace{-3mm} \begin{array}{l}\textsf{s}_{2,1}\oplus2\textsf{n}_{1,1},\text{ or}\\ \textsf{s}_{5,14}\text{ if }A_2=0\end{array}$\\
\vspace{1mm}

5 &$\begin{array}{l}A^1= F(x_3),\ A^2=G(x_3),\\A^0=R(x_3)+\omega_1x_1A^1+\omega_2x_2A^2\\-\frac{\omega_1^2x_1^2}2-
\frac{\omega_2^2x_2^2}2\end{array}$&$\begin{array}{l} B^+_1(\omega_1),\  B^-_1(\omega_1),\\ B^+_2(\omega_2),\ B^-_2(\omega_2) \end{array}$&$\hspace{-3mm} \begin{array}{l}\textsf{s}_{6,162} \end{array}$\\

\vspace{1mm}

6\hspace{1mm}&$\begin{array}{l}A^1=F(x_3)-\alpha x_2,\  A^2=G(x_3)+\alpha x_1,\\A^0=R(x_3)-\frac{\omega^2\tilde r^2}2+2\alpha\omega x_1x_2
\end{array}$&$\begin{array}{l} B^+_1(\omega)-\alpha \text{e}^{\omega t}x_2,\\ B^-_2(\omega)+\alpha \text{e}^{-\omega  t}x_1\end{array}$&$\hspace{-3mm} \begin{array}{l}\textsf{s}_{4,6}  \end{array}$\\

7$^\star$\hspace{1mm} \vspace{1mm}&$\begin{array}{l}A^1=-\alpha x_2,\  A^2=\alpha x_1,\\A^0=-\frac{\omega^2x_3^2}2\end{array}$
&$\begin{array}{l} P_1-\alpha x_2,\  P_2+\alpha x_1,\\ B^+_3(\omega), \ B^-_3(\omega) \ L_3, \end{array}$&$\textsf{s}_{7,2}$\\

8$^\star$\hspace{1mm} \vspace{2mm}&$\begin{array}{l}A^1=-\alpha x_2,\  A^2=\alpha x_1,\\A^0=-\frac{\omega^2\tilde r^2}2+2\alpha\omega x_1x_2
\end{array}$&$\begin{array}{l} B^+_1(\omega)-\alpha \text{e}^{\omega t}x_2, \ G_3 \\B^-_2(\omega)+\alpha \text{e}^{-\omega  t}x_1,\ P_3\end{array}$&$\hspace{-3mm} \begin{array}{l}\textsf{s}_{6,160}  \end{array}$\\

9\hspace{1mm} \vspace{2mm}&$\begin{array}{l}A^1=-\alpha x_2,\  A^2=\alpha x_1,\\A^0=2\alpha\omega x_1x_2-\frac{\omega^2\tilde r^2}2-\frac{\omega_3^2x_3^2}2
\end{array}$&$\begin{array}{l} B^+_1(\omega)-\alpha \text{e}^{\omega t}x_2,\\ B^-_2(\omega)+\alpha \text{e}^{-\omega  t}x_1,\\B^+_3(\omega_3), \ B^-_3(\omega_3)\end{array}$&$\hspace{-3mm} \begin{array}{l}\textsf{s}_{6,162}  \end{array}$\\

10$\hspace{1mm} $
&  \vspace{2mm}$ \begin{array}{l}A^1=\p_1\tilde F(\theta,\varphi)+\p_2G(\theta,\varphi),\\ A^2=\p_2(\tilde F(\theta,\varphi)-\p_1G(\theta,\varphi),\\A^0=-\frac{\omega^2 r^2}2+\frac1{r^2}R(\theta,\varphi)\\+\omega(\p_\varphi G(\theta,\varphi)-\frac12\sin(2\theta)\p_\theta F(\theta,\varphi))\end{array}$
 \vspace{0mm}
 &$\begin{array}{l}A^{+}(\omega) \end{array}$&$\hspace{-3mm} \begin{array}{l} \textsf{s}_{2,1}\oplus\textsf{n}_{1,1}\end{array}$\\

11$\hspace{1mm} $
&  \vspace{2mm}$ \begin{array}{l}A^1=\p_1\tilde F(\theta)+\p_2G(\theta),\\ A^2=\p_2\tilde F(\theta)-\p_1G(\theta),\\A^0=-\frac{\omega^2 r^2}2+\frac1{r^2}R(\theta)\\-\frac{\omega}2\sin(2\theta)\p_\theta F(\theta)\end{array}$
 \vspace{0mm}
 &$\begin{array}{l}A^{+}(\omega),\ L_3 \end{array}$&$\hspace{-3mm} \begin{array}{l} \textsf{s}_{2,1}\oplus2\textsf{n}_{1,1}\end{array}$\\

12&\vspace{2mm}$\begin{array}{l}A^1=x_3\p_1\frac {F({\varphi})}{\tilde r},\\  A^2= x_3 \p_2\frac {F({\varphi})}{\tilde r},\\A^0= \frac1{\tilde r^2}R(\varphi)-\frac{\omega^2r^2}2+\frac{\omega x_3 F(\varphi)}{\tilde r}\end{array}$&$\begin{array}{l}A^{+}(\omega),\\B^+_3(\omega)-\frac{\text{e}^{\omega t} {F({\varphi})}}{\tilde r} \end{array}
$&$\hspace{-3mm} \begin{array}{l}\textsf{s}_{3,1}\oplus\textsf{n}_{1,1}\end{array}$\\
\vspace{0mm}

13&\vspace{2mm}$\begin{array}{l}A^1=\frac {x_1x_3}{\tilde r^3},\  A^2= \frac {x_2x_3}{\tilde r^3},\\A^0= \frac\nu{\tilde r^2}-\frac{\omega^2r^2}2- \frac{ \omega x_3} {\tilde r}\end{array}$&$\begin{array}{l}A^{+}(\omega),\ L_3,\\B^+_3(\omega)+\frac{\text{e}^{\omega t}}{\tilde r} \end{array}
$&$\hspace{-3mm} \begin{array}{l}\textsf{s}_{3,1}\oplus2\textsf{n}_{1,1}\end{array}$\\

14&\vspace{1mm}$\begin{array}{l}A^1=\frac {x_3 }{x_1^2},\  A^2= 0,\\A^0= -\frac{\omega^2r^2}2-\frac{\omega x_3}{x_1} \end{array}$&$\begin{array}{l}A^{+}(\omega),\ B_2^+,\ B_2^-\\B^+_3(\omega)+\frac{\text{e}^{\omega t}}{ x_1} \end{array}
$&$\hspace{-3mm} \begin{array}{l}\textsf{s}_{6,124}\end{array}$\\

\hline\hline

\end{tabular}

\vspace{0mm}

\section{Discussion}

We have classified
the continuous point symmetries  admitted  by the SE for a charged particle interacting with electric and magnetic fields. The classification results are presented in four tables which include qualitatively different systems.

Symmetries presented in Table 1  either belong or can be reduced to symmetries of  the   free SE. The reduction is needed only for the systems presented in Items 1, 4, 5 and can be made using the gauge transformations. It is possible to nullify the external electric field, or  magnetic field, or both of them, and the mentioned symmetries still would be valid. In other words, the external fields presented there cause the reduction of symmetries of the free SE but keep a part of them.

Symmetries collected in Table 2 do not belong to the symmetries of the free SE. If we nullify the  scalar potentials these symmetries are still valid. The presence of the vector potentials is essential since they are constituent parts of the mentioned symmetries, which in fact are induced by the external magnetic fields.

The systems fixed in Table 3 also admit symmetries which are not valid for the free SE. The vector potentials presented there can be nullified without  reduction or extension of the admitted symmetries. The scalar potentials should be non-trivial since in fact they generates the mentioned symmetries. Notice that the symmetries presented in Items 6-14 belong to subalgebras of the symmetry algebra of repulsive oscillator.

The most specific systems can be found in Table 4. In order to  the symmetries presented here be valid, both the vector and scalar potentials should be non-trivial.

  In accordance with the above,  Tables 1 and 3 in particular include the classification of symmetries of SEs with scalar potentials since the vector potentials presented there can be removed without changing the symmetry groups of the corresponding equations. In this way we recover the Boyer classification \cite{Boy} or, more exactly, its corrected version presented in \cite{Nuca}.   On the other hand, Tables 1 and 2 implicitly include the classification of symmetries of SEs describing particles interacting with  purely magnetic field since the scalar potentials presented there  can be  nullified without changing the presented symmetries. And only Table 4 includes the systems with non-separable interactions with superpositions of the electric and magnetic fields. Namely, neither scalar nor vector potentials cannot be nullified without changing the symmetries presented there, and these symmetries are not valid for the free SE.

In the classification tables the systems admitting additional equivalence transformations are clearly indicated. The invariance algebras of dimension $d\leq 6$ are specified  using notations proposed in \cite{snob}. More extended solvable  invariance algebras are notified by the symbols  $\textsf{s}_{n,m}$  where the first subindex $n>6$ indicates the algebra dimension and the second one is a conventional number of the algebra. Since solvable algebras of such dimensions are not classified jet, everything we could do was to present the nontrivial commutation relations for their basis elements.

The presented specification of the symmetry algebras might be useful for potential readers. Various fine properties  of these algebras including explicit forms of the related Casimir operators can be found in  \cite{snob}.

The majority of the scalar potentials presented  in Tables 3 and 4 include the terms which are quadratic in space variables. Moreover, in Table 3 we can treat these terms as either repulsive or harmonic oscillator since parameters $\omega$ and $\omega_a$  can be either real or purely imaginary. Formally speaking the same is true for the potentials presented in Table 4, but there is a specific point missing for the systems presented in Table 3. Namely, for $\omega$ imaginary the related potentials became complex and so they cannot be used in frames of the standard quantum mechanics. However they have perfect perspectives in the "PT-symmetric quantum mechanics" \cite{benya} since many of them have good properties with respect to the space inversion and all of them are compatible with this property provided it would be requested additionally.

Some of the presented systems have rather extended symmetries.
In many cases just  such  systems  have good application perspectives.  A well known example is the Landau system which describes the interaction of a charged particle with the constant and homogeneous magnetic field and accepts the seven dimensional Lie algebra including four integrals of motion. Just this system is presented in Item 8 of Table 2. Let us note that this system admits  interesting generalizations to the case of the superposition of the constant magnetic and hyperbolic  scalar potential,  see Item 5, 6 and 7 of Table 4. The generalizations keep the symmetry of the Landau system  w.r.t. the six dimensional Lie algebras, moreover, for the system presented in Item 5 the corresponding symmetry algebra is seven dimensional.

Quantum mechanical systems with extended Lie symmetries in many cases are exactly solvable. This is the case also for systems with position dependent mass. Namely, a symmetry with respect to six dimension Lie algebra guarantees their exact solvability \cite{N22}.  It looks like that the same is true for the equations  discussed in the present paper, but this statement needs a rather sophisticated  verification.

The  present paper completes and finishes   the group classification of SEs with time independent potentials which started with paper \cite{Boy} where the case of scalar potentials was considered and continued in papers \cite{NZ2, NN} where the equations with position dependent mass were classified and paper \cite{NNN} where symmetries of SEs with matrix potentials had been found.

The presented classification is complete. Namely,  all inequivalent systems and the corresponding symmetry algebras are presented. The equivalence groupoid for the considered class of equations is presented  also. By definition it includes all symmetry algebras (in our case they are fixed in Tables 1--4) and additional equivalence transformations which do not belong to these groups and are described in Section 5. In particular we find  a "maravillosas" form (\ref{t8}),  (\ref{t9}) of equivalence transformations which are accepted by all systems invariant with respect to the scaling of independent variables.

However, still  there are challenges for experts in group theoretical methods in physics connected with SE. First an open research field includes  symmetries of SEs with time dependent potentials. In fact the only known complete result in this field  is the classification of (1+1)--dimensional SEs  with (complex) scalar  potentials \cite{popa1}. Such systems of more high dimension and especially ones including vector potentials are still waiting their researchers. An open problem is the classification of symmetries of Schr\"odinger-Pauli equations for charged particles. But maybe a more big challenge is the classification of higher symmetries of 3d Schr\"odinger equations with position dependent mass, and just this business is in progress in our department.

 Let us note that symmetries of  SEs with scalar and vector potentials were studied also in papers \cite{bol} and \cite{maga}. In \cite{bol} the class of the external fields is strongly restricted by the supposition that they are invariant with respect to three dimension subalgebras of algebra $\textsf{e(3)}$ and in fact only symmetries of such fields (which potentially may be useful in searching for SE symmetries) are discussed.  In \cite{maga} symmetries of  stationary Schr\"odinger  equations are classified. Such equations are particular cases of the time dependent SEs considered in the present  paper, and  it is possible to compare the obtained results which appears to be  are in a good accordance. However, the results presented in \cite{maga}   have a rather cumbersome  form which can be essentially simplified using    gauge transformations.

For group classification of nonlinear and generalized nonlinear SEs see \cite{pop} and  \cite{Nn2, Nn1, popa, cina}.


\end{document}